\documentclass[epj]{svjour}
\usepackage{verbatim} 

\usepackage{graphicx}
\usepackage{dcolumn}

\usepackage{amstext,amsmath,amssymb,amsfonts}
\usepackage{bm}

\usepackage{txfonts}
\usepackage{mathtools}

\usepackage{hyperref}

\newcommand{\bra}[1]{\ensuremath{\langle #1 \vert}}
\newcommand{\ket}[1]{\ensuremath{\vert #1  \rangle}}

\renewcommand{\b}[1]{\ensuremath{\mathbf{#1}}}

\renewcommand{\l}{\ensuremath{\lambda}}

\newcommand{\Tr}{\ensuremath{\text{Tr}}}

\renewcommand{\star}{\ensuremath{\text{o}}}

\renewcommand{\d}{\ensuremath{\text{d}}}

\DeclareMathOperator{\arccot}{arccot}

\newcommand{\isEquivTo}[1]{\underset{#1}{\sim}}

\renewcommand{\i}{\ensuremath{\text{i}}}

\begin{document}

\title{Semiclassical approximations of photoabsorption cross sections beyond the continuum threshold}
\dedication{Dedicated to the memory of Peter Schuck}

\author{Julien Toulouse}
\mail{toulouse@lct.jussieu.fr}
\institute{Laboratoire de Chimie Théorique, Sorbonne Université and CNRS, F-75005 Paris, France \and 
Institut Universitaire de France, F-75005 Paris, France}

\date{April 22, 2023}

\abstract{
We develop semiclassical approximations for calculating photoabsorption cross sections beyond the continuum threshold in quantum many-body systems. These approximations use the fully quantum-mechanical Wigner function of the ground state and semiclassical expansions only for the part of the cross section depending on the continuum states, thus avoiding the difficult explicit calculation of the continuum states. Even though the approach is general, we test it in electronic-structure theory for the photoionization cross sections of the hydrogen and helium atoms. The results suggest that these semiclassical approximations can be used to obtain good estimates of cross sections at high energy.
}

\maketitle

\section{Introduction}

In many-body quantum systems, the calculation of properties involving continuum states constitutes a challenge for computational methods. The simplest example is perhaps given by the photoabsorption cross section beyond the continuum threshold, corresponding to transitions of the system from a bound state to continuum states induced by the absorption of a photon (see, e.g., Refs.~\cite{RinSch-BOOK-04,KrePabSan-AJP-14}). In the context of the electronic-structure theory of atoms and molecules, this property is also known as the photoionization cross section since it corresponds to the ionization of the system by ejection of one or more electrons into the continuum (see, e.g., Ref.~\cite{GreDec-CCR-05}). Hence, calculations of photoabsorption/photoionization cross sections require an appropriate description of excitations to continuum states and are usually performed with quite sophisticated and computationally expensive approaches, i.e. using extended basis sets such as B-spline basis sets~\cite{DecLisVen-JPB-94,Mar-JPB-99,BacCorDecHanMar-RPP-01} with continuum boundary conditions~\cite{BurNobBur-INC-07,ZatBar-JPB-13,SchZapLevCanLupTou-JCP-22} or using techniques involving the complex-energy plane such as complex scaling~\cite{Chu-PRL-97,Chu-RPC-04}, analytical continuation~\cite{TenCorRocNas-PTCP-21}, or integral transforms~\cite{EfrLeiOrlBar-JPG-07}.

In this work, as a possible alternative to these involved fully quantum mechanical calculations, we develop semiclassical approximations for calculating photoabsorption cross sections, based on the Wigner phase-space formulation of quantum mechanics (see, e.g., Refs.~\cite{HilOcoScuWig-PR-84,OsbMol-AP-95,RinSch-BOOK-04,WeiFer-APR-18}) or also known as deformation quantization~\cite{Zac-IJMP-02,HanWalWyn-EJP-04}. Wigner-based semiclassical approximations have been used in nuclear physics to calculate various quantities (see, e.g., Refs.~\cite{RinSch-BOOK-04}). In particular, the full semiclassical Wigner-Kirkwood expansion (in powers of the reduced Planck constant $\hbar$) of the linear-response function has been determined~\cite{ChaSch-PRA-88,SchHasJaeGreRemSebSur-PPNP-89}. Here, we consider semiclassical approximations of the photoabsorption cross section using the fully quantum-mechanical Wigner function of the ground state and semiclassical expansions only for the part depending on the continuum states. This is motivated by the fact that semiclassical expansions are expected to work better for continuum states than for the ground state. Similar semiclassical approximations have been used in molecular physics to calculate photodissociation cross sections~\cite{Hel-JCP-78,SheWal-JCP-83,HupEck-JCP-99,McqAbrBru-JCP-03} (see, also, Refs.~\cite{HupEck-PRA-98,Seg-JOB-03}), but to the best our knowledge this type of semiclassical approximations have never been developed for photoabsorption cross sections (see, however, Ref.~\cite{BanPelWeiMayFil-PRB-22} for a Wigner-based approach of light absorption in solids). As an illustration, we test this approach in electronic-structure theory for the photoionization cross sections of the hydrogen and helium atoms, but it can be a priori applied to the photoabsorption cross sections appearing in other fields such as nuclear physics.

The paper is organized as followed. In Section~\ref{sec:general}, we lay down the general theory for semiclassical approximations of photoabsorption cross sections for an arbitrary $N$-particle system. In Section~\ref{sec:spherical}, we treat the case of one-particle systems with spherical ground states. In Section~\ref{sec:hydrogen}, we work out the specific case of hydrogen-like atoms with a Coulomb potential, and give results for the hydrogen atom. In Section~\ref{sec:helium}, we treat the case of helium-like atoms, and give results for the helium atom. Finally, Section~\ref{sec:conclusion} contains conclusions and future directions. Hartree atomic units (a.u.), in which $\hbar=m=e=1/(4\pi\epsilon_0)=1$, are used throughout this work.

This work was started together with the late Peter Schuck who developed semiclassical approximations in nuclear physics and was eager to extend them to other fields. The present author is thus very much indebted to Peter Schuck for having introduced him to these Wigner-based semiclassical techniques and guided him through the early stages of the present work. The paper is thus dedicated to his memory.

\section{General theory for $N$-particle systems}
\label{sec:general}

\subsection{Photoabsorption cross section in terms of Wigner transforms}

We consider a non-relativistic $N$-particle Hamiltonian, 
\begin{eqnarray}
\hat{H} = \frac{\hat{\b{p}}^2}{2} + \hat{V},
\label{H}
\end{eqnarray}
where $\hat{\b{p}}=(\hat{\b{p}}_1,\hat{\b{p}}_2,\cdots,\hat{\b{p}}_N)$ collects all momentum operators of the individual particles and $\hat{V}$ is a potential-energy operator, with eigenstates $\{ \ket{\Psi_n} \}_{n\in \mathbb{N}}$ and eigenvalues $\{ E_n \}_{n\in \mathbb{N}}$
\begin{eqnarray}
\hat{H} \ket{\Psi_n} = E_n \ket{\Psi_n}.
\label{HPsiEPsi}
\end{eqnarray}
Denoting by $E_\text{thres}$ the continuum threshold energy, the eigenstates with $E_n < E_\text{thres}$ are bound states and the eigenstates with $E_n \geq E_\text{thres}$ are continuum states assumed to be discretized for simplicity (e.g., obtained by putting the system in a large finite box with periodic boundary conditions), so that $\{ \ket{\Psi_n} \}_{n\in \mathbb{N}}$ forms a discrete complete orthonormal basis of the Hilbert space. The linear-response photoabsorption cross section, corresponding to transitions between the ground state $\ket{\Psi_0}$ and the excited states $\ket{\Psi_n}$, in the velocity-gauge electric-dipole approximation at frequency $\omega$ is defined as
\begin{eqnarray}
\sigma(\omega) = \frac{4\pi^2}{3c\omega} \sum_{\mu \in\{x,y,z\}} \sum_{n=0}^\infty   |\bra{\Psi_0} \hat{P}_\mu \ket{\Psi_{n}}|^2 \; \delta(\omega - (E_n - E_0)),
\nonumber\\
\label{Vomega}
\end{eqnarray}
where $c=137.036$ a.u. is the speed of light, $\hat{P}_\mu = \sum_{i=1}^N \hat{p}_{i,\mu}$ is the Cartesian $\mu$-component of the total momentum (or velocity) operator, and $\delta$ is the Dirac delta function. We are interested in the cross section beyond the continuum threshold, i.e. $\omega \geq E_\text{thres} - E_0$. The cross section can be rewritten as
\begin{eqnarray}
\sigma(\omega) &=& \frac{4\pi^2}{3c\omega} \sum_{\mu \in\{x,y,z\}} \sum_{n=0}^\infty \bra{\Psi_0} \hat{P}_\mu \ket{\Psi_n} \bra{\Psi_n} \delta (\omega + E_0 - \hat{H} ) \hat{P}_\mu \ket{\Psi_0}
\nonumber\\
          &=& \frac{4\pi^2}{3c\omega} \sum_{\mu \in\{x,y,z\}} \bra{\Psi_0} \hat{P}_\mu \delta (\omega + E_0 - \hat{H} ) \hat{P}_\mu \ket{\Psi_0}
\nonumber\\
          &=& \frac{4\pi^2}{3c\omega} \Tr[ \hat{B} \; \hat{\rho}_0],
\label{Vomegatrace}
\end{eqnarray}
where we have used the Schr\"odinger equation [Eq.~(\ref{HPsiEPsi})] and the completeness relation $\sum_{n=0}^\infty \ket{\Psi_n} \bra{\Psi_n} = \hat{1}$, and we have introduced the operator 
\begin{eqnarray}
\hat{B}= \sum_{\mu \in\{x,y,z\}} \hat{P}_\mu \hat{A} \hat{P}_\mu,
\label{Bop}
\end{eqnarray}
with the spectral-density operator $\hat{A} = \delta (\omega + E_0 - \hat{H})$ and the ground-state density-matrix operator
\begin{eqnarray}
\hat{\rho}_0 = \ket{\Psi_0} \bra{\Psi_0}.
\label{}
\end{eqnarray}

In the position representation, the cross section takes the form
\begin{eqnarray}
\sigma(\omega) = \frac{4\pi^2}{3c\omega} \int_{\mathbb{R}^{6N}} \d \b{r} \d \b{r}' B (\b{r},\b{r}') \rho_0 (\b{r}',\b{r}), 
\label{}
\end{eqnarray}
where $\b{r}=(\b{r}_1,\b{r}_2,\cdots,\b{r}_N) \in \mathbb{R}^{3N}$ and $\b{r}'=(\b{r}_1',\b{r}_2',\cdots,\b{r}_N') \in \mathbb{R}^{3N}$ are position vectors of the $N$ particles, and $B(\b{r},\b{r}') = \bra{\b{r}} \hat{B} \ket{\b{r}'}$ and $\rho_0 (\b{r}',\b{r}) =  \bra{\b{r}'} \hat{\rho}_0 \ket{\b{r}}=\Psi_0^{}(\b{r}') \Psi_0^*(\b{r})$. We now introduce the Wigner (or Weyl) transforms/representations of the operators $\hat{B}$ and $\hat{\rho}_0$ (see, e.g., Refs.~\cite{RinSch-BOOK-04,Cas-AJP-08})
\begin{eqnarray}
[\hat{B}]_\text{W}(\b{q},\b{p}) \equiv B_\text{W}(\b{q},\b{p}) =  \int_{\mathbb{R}^{3N}} \d \b{s} \; e^{-i \b{p} \cdot \b{s}} \bra{\b{q}+\b{s}/2} \hat{B} \ket{\b{q}-\b{s}/2},
\label{BWdef}
\end{eqnarray}
\begin{eqnarray}
[\hat{\rho}_0]_\text{W} (\b{q},\b{p}) \equiv \rho_{0,\text{W}}(\b{q},\b{p}) =  \int_{\mathbb{R}^{3N}} \d \b{s} \; e^{-i \b{p} \cdot \b{s}} \bra{\b{q}+\b{s}/2} \hat{\rho}_0 \ket{\b{q}-\b{s}/2}, \;\;
\label{rho0W}
\end{eqnarray}
where $\b{q}=(\b{r}+\b{r}')/2 \in \mathbb{R}^{3N}$ is the average position vector, $\b{s}=\b{r}-\b{r}'\in \mathbb{R}^{3N}$ is the relative position vector, and $\b{p}=(\b{p}_1,\b{p}_2,\cdots,\b{p}_N) \in \mathbb{R}^{3N}$ is the conjugate momentum vector of $\b{s}$. The Wigner transformation preserves the trace of a product of operators, so we have 
\begin{eqnarray}
\sigma(\omega) = \frac{4\pi^2}{3c\omega} \int_{\mathbb{R}^{6N}} \frac{\d \b{q} \d \b{p}}{(2\pi)^{3N}} \; B_\text{W}(\b{q},\b{p}) \rho_{0,\text{W}}(\b{q},\b{p}).
\label{vomegaintdqdp}
\end{eqnarray}
We have thus put the photoabsorption cross section in the form of a phase-space integral. So far, everything is exact. We will assume that we know the Wigner function of the ground state $\rho_{0,\text{W}}(\b{q},\b{p})$, and we will now use a semiclassical expansion approximation for $B_\text{W}(\b{q},\b{p})$.

\subsection{Semiclassical expansion approximation}

A convenient formula for calculating Wigner transforms and their semiclassical expansions is the following expression for the Wigner transform of the product of two operators $\hat{C}$ and $\hat{D}$, also known as Groenewold's formula or Moyal product or star product (see Ref.~\cite{RinSch-BOOK-04}),
\begin{eqnarray}
[ \hat{C} \hat{D} ]_\text{W}(\b{q},\b{p}) = C_\text{W}(\b{q},\b{p}) e^{(i\hbar/2) \overset\leftrightarrow{\Lambda}} D_\text{W}(\b{q},\b{p}),
\label{CDW}
\end{eqnarray}
where $\overset\leftrightarrow{\Lambda} = \overset\leftarrow{\nabla}_{\b{q}} \cdot \overset\rightarrow{\nabla}_{\b{p}} - \overset\leftarrow{\nabla}_{\b{p}} \cdot \overset\rightarrow{\nabla}_{\b{q}}$ is the Poisson bracket differential operator (the arrows indicate on which side act the nabla operators) and the reduced Planck constant $\hbar=1$ a.u. is kept to keep track of orders in $\hbar$. Expanding Eq.~(\ref{CDW}) in powers of $\hbar$ generates a semiclassical expansion.

By repeatedly applying this formula, we can write $B_\text{W}(\b{q},\b{p})$ as
\begin{eqnarray}
B_\text{W}(\b{q},\b{p}) 
&=& \sum_{\mu\in\{x,y,z\}} P_\mu  A_\text{W} P_\mu - \frac{\hbar^2}{4} P_\mu \overset\leftrightarrow{\Lambda} (A_\text{W} \overset\leftrightarrow{\Lambda}  P_\mu),
\label{BWexpand}
\end{eqnarray}
where $A_\text{W} \equiv A_\text{W}(\b{q},\b{p})$ is the Wigner transform of the operator $\hat{A}$. We have used the fact that the Wigner transform of total momentum operator $\hat{P}_\mu$ is the total momentum variable $P_\mu= \sum_{i=1}^N p_{i,\mu}$, i.e. $[\hat{P}_\mu]_\text{W} = P_\mu$, and we have used the fact that the antisymmetry of the Poisson bracket differential operator $\overset\leftrightarrow{\Lambda}$ implies that $P_\mu \overset\leftrightarrow{\Lambda} A_\text{W} = - A_\text{W} \overset\leftrightarrow{\Lambda}  P_\mu$ and $P_\mu \overset\leftrightarrow{\Lambda} P_\mu=0$. Note that there is no higher-order terms in Eq.~(\ref{BWexpand}) since acting twice with $\overset\leftrightarrow{\Lambda}$ on $P_\mu$ gives zero. The last term in Eq.~(\ref{BWexpand}) is
\begin{eqnarray}
P_\mu \overset\leftrightarrow{\Lambda} (A_\text{W} \overset\leftrightarrow{\Lambda}  P_\mu) 
&=& - {\nabla}_{\b{p}} P_\mu  \cdot {\nabla}_{\b{q}} ( {\nabla}_{\b{q}} A_\text{W} \cdot {\nabla}_{\b{p}} P_\mu)
\nonumber\\
&=& - \sum_{i=1}^N \sum_{j=1}^N \frac{\partial^2 A_\text{W}}{\partial q_{i,\mu} \partial q_{j,\mu}}.
\label{}
\end{eqnarray}
We thus obtain the following exact expression for $B_\text{W}(\b{q},\b{p})$
\begin{eqnarray}
B_\text{W}(\b{q},\b{p}) &=& \sum_{i=1}^N \sum_{j=1}^N \sum_{\mu\in\{x,y,z\}} \left[ p_{i,\mu} p_{j,\mu} A_\text{W}(\b{q},\b{p}) + \frac{\hbar^2}{4} \frac{\partial^2 A_\text{W}(\b{q},\b{p})}{\partial q_{i,\mu} \partial q_{j,\mu}} \right],
\nonumber\\
&=& \b{P}^2 A_\text{W}(\b{q},\b{p}) + \frac{\hbar^2}{4} \b{D}^2 A_\text{W}(\b{q},\b{p}),
\label{}
\end{eqnarray}
where we have introduced the total momentum vector $\b{P} = \sum_{i=1}^N \b{p}_{i}$ and the differential operator $\b{D} = \sum_{i=1}^N {\nabla}_{\b{q}_i}$.

It remains to find an expression for $A_\text{W}(\b{q},\b{p})$, i.e. the Wigner transform of the operator $\hat{A} = \delta (\omega + E_0 - \hat{H})$. This Wigner transform cannot be calculated exactly but we can write its second-order semiclassical expansion
\begin{eqnarray}
A_\text{W}(\b{q},\b{p}) = A_\text{W}^{(0)}(\b{q},\b{p}) + \hbar^2 A_\text{W}^{(2)}(\b{q},\b{p}) + O(\hbar^4),
\label{}
\end{eqnarray}
where the zeroth-order term is
\begin{eqnarray}
A_\text{W}^{(0)}(\b{q},\b{p})  = \delta \!\left(\omega + E_0 - H(\b{q},\b{p})\right),
\label{AW0}
\end{eqnarray}
where $H(\b{q},\b{p}) = \b{p}^2/2 + V(\b{q})$ is the classical Hamiltonian, and the second-order correction is obtained from Eq.~(13.43) of Ref.~\cite{RinSch-BOOK-04} (by differentiating with respect to $\l_\text{sc}$ and correcting the minus sign in front of $\delta''$ into a plus sign) (see also Refs.~\cite{DurSchTre-JP-84,BalJen-PR-84,CenVinDurSchVon-AP-98,VinSchFarCen-PRC-03})
\begin{eqnarray}
A_\text{W}^{(2)}(\b{q},\b{p}) = \frac{1}{8} \Biggl[ -\nabla_\b{q}^2 V(\b{q})  \;\delta''\!\left(\omega + E_0 - H(\b{q},\b{p})\right) \phantom{xxxxxxxx}
\nonumber\\
+ \frac{1}{3} \left( (\nabla_\b{q} V(\b{q}))^2 + (\b{p} \cdot \nabla_\b{q})^2 V(\b{q}) \right) \; \delta'''\!\left(\omega + E_0 -H(\b{q},\b{p})\right) \Biggl]. \;\;
\label{AW2}
\end{eqnarray}
We thus obtain the second-order semiclassical expansion for $B_\text{W}(\b{q},\b{p})$
\begin{eqnarray}
B_\text{W}(\b{q},\b{p}) = B_\text{W}^{(0)}(\b{q},\b{p}) + \hbar^2 B_\text{W}^{(2)}(\b{q},\b{p}) + O(\hbar^4),
\label{}
\end{eqnarray}
with
\begin{eqnarray}
B_\text{W}^{(0)}(\b{q},\b{p}) = \b{P}^2 A_\text{W}^{(0)}(\b{q},\b{p}),
\label{BW0}
\end{eqnarray}
and
\begin{eqnarray}
B_\text{W}^{(2)}(\b{q},\b{p}) &=& \frac{1}{4} \b{D}^2 A_\text{W}^{(0)}(\b{q},\b{p}) + \b{P}^2 A_\text{W}^{(2)}(\b{q},\b{p}).
\label{BW2}
\end{eqnarray}
The expression of $\b{D}^2 A_\text{W}^{(0) }(\b{q},\b{p})$ is obtained directly from Eq.~(\ref{AW0})
\begin{eqnarray}
\b{D}^2 A_\text{W}^{(0)}(\b{q},\b{p})&=& - \left( \b{D}^2 V(\b{q}) \right) \;\delta'\!\left(\omega + E_0 - H(\b{q},\b{p})\right) 
\nonumber\\
&&+ \left( \b{D} V(\b{q}) \right)^2 \; \delta''\!\left(\omega + E_0 -H(\b{q},\b{p})\right).
\label{dAdqdq}
\end{eqnarray}
Finally, we obtain the second-order semiclassical expansion of the photoabsorption cross section [Eq.~(\ref{vomegaintdqdp})]
\begin{eqnarray}
\sigma(\omega) = \sigma^{(0)}(\omega) + \sigma^{(2)}(\omega) + \cdots,
\label{sigmaomegaexpand}
\end{eqnarray}
with the zeroth-order cross section
\begin{eqnarray}
\sigma^{(0)}(\omega) &=& \frac{4\pi^2}{3c\omega} \int_{\mathbb{R}^{6N}} \frac{\d \b{q} \d \b{p}}{(2\pi)^{3N}} \; B_\text{W}^{(0)}(\b{q},\b{p}) \rho_{0,\text{W}}(\b{q},\b{p})
\nonumber\\
&=& \frac{4\pi^2}{3c\omega} \int_{\mathbb{R}^{6N}} \frac{\d \b{q} \d \b{p}}{(2\pi)^{3N}} \; \b{P}^2 \delta \!\left(\omega + E_0 - H(\b{q},\b{p})\right) \rho_{0,\text{W}}(\b{q},\b{p}),
\nonumber\\
\label{sigma0omega}
\end{eqnarray}
and the second-order correction
\begin{eqnarray}
\sigma^{(2)}(\omega) = \frac{4\pi^2}{3c\omega} \int_{\mathbb{R}^{6N}} \frac{\d \b{q} \d \b{p}}{(2\pi)^{3N}} \; B_\text{W}^{(2)}(\b{q},\b{p}) \rho_{0,\text{W}}(\b{q},\b{p}).
\label{sigma2omega}
\end{eqnarray}
We will write the latter term as a sum of three contributions
\begin{eqnarray}
\sigma^{(2)}(\omega) = \sigma^{(2a)}(\omega) + \sigma^{(2b)}(\omega) + \sigma^{(2c)}(\omega),
\label{sigma2omega}
\end{eqnarray}
where $\sigma^{(2a)}(\omega)$ is the contribution coming from the $\delta'$ function in Eq.~(\ref{dAdqdq})
\begin{eqnarray}
\sigma^{(2a)}(\omega) &=& -\frac{\pi^2}{3c\omega} \int_{\mathbb{R}^{6N}} \frac{\d \b{q} \d \b{p}}{(2\pi)^{3N}} \; \left( \b{D}^2 V(\b{q}) \right) \phantom{xxxxx}
\nonumber\\
&&\times\; \delta'\!\left(\omega + E_0 - H(\b{q},\b{p})\right) \rho_{0,\text{W}}(\b{q},\b{p}),
\label{sigma2aomega}
\end{eqnarray}
$\sigma^{(2b)}(\omega)$ is the contribution coming from the $\delta''$ function in Eqs.~(\ref{AW2}) and~(\ref{dAdqdq})
\begin{eqnarray}
\sigma^{(2b)}(\omega) &=& \frac{\pi^2}{3c\omega} \int_{\mathbb{R}^{6N}} \frac{\d \b{q} \d \b{p}}{(2\pi)^{3N}} \; \left[  \left( \b{D} V(\b{q}) \right)^2 - \frac{\b{P}^2}{2} \nabla_\b{q}^2 V(\b{q}) \right]
\nonumber\\
&&\times\; \delta''\!\left(\omega + E_0 -H(\b{q},\b{p})\right) \rho_{0,\text{W}}(\b{q},\b{p}),
\label{sigma2bomega}
\end{eqnarray}
and $\sigma^{(2c)}(\omega)$ is the contribution coming from the $\delta'''$ function in Eq.~(\ref{AW2})
\begin{eqnarray}
\sigma^{(2c)}(\omega) &=& \frac{\pi^2}{18c\omega} \int_{\mathbb{R}^{6N}} \frac{\d \b{q} \d \b{p}}{(2\pi)^{3N}} \; \b{P}^2 \left( (\nabla_\b{q} V(\b{q}))^2 + (\b{p} \cdot \nabla_\b{q})^2 V(\b{q}) \right) 
\nonumber\\
&& \times\; \delta'''\!\left(\omega + E_0 -H(\b{q},\b{p})\right) \rho_{0,\text{W}}(\b{q},\b{p}). \;\;\; 
\label{sigma2comega}
\end{eqnarray}

We have thus arrived at an approximation to the photoabsorption cross section that only requires to know the ground-state Wigner function $\rho_{0,\text{W}}(\b{q},\b{p})$ but does not require the calculation of the continuum states. Note that Eq.~(\ref{sigmaomegaexpand}) is not a full expansion in powers of $\hbar$ since we do not expand $\rho_{0,\text{W}}(\b{q},\b{p})$ in powers of $\hbar$.

\section{Theory for one-particle systems with spherical ground states}
\label{sec:spherical}

We now apply the general theory of the previous section to the case of one-particle systems ($N=1$) with spherical ground states. The phase-space variables are now $\b{q} \equiv \b{q}_1 \in \mathbb{R}^3$ and $\b{p} \equiv \b{p}_1 \in \mathbb{R}^3$, and the classical Hamiltonian is
\begin{eqnarray}
H(\b{q},\b{p}) =\frac{p^2}{2} + V(q),
\end{eqnarray}
where $p=||\b{p}||$ and $q=||\b{q}||$, and $V(q)$ is a central potential. This case encompasses not only one-electron atoms but also many-electron atoms within a mean-field approximation with a spherical local potential such as Kohn-Sham density-functional theory. The ground-state Wigner function then depends only on $q$, $p$, and $\b{q}\cdot\b{p}$, i.e. $\rho_{0,\text{W}}(\b{q},\b{p}) \equiv \rho_{0,\text{W}}(q,p,\b{q}\cdot\b{p})$.

\subsection{Zeroth-order semiclassical approximation}

The zeroth-order photoabsorption cross section [Eq.~(\ref{sigma0omega})] simplifies to
\begin{eqnarray}
\sigma^{(0)}(\omega) &=& \frac{4\pi^2}{3c\omega} \int_{\mathbb{R}^{6}} \frac{\d \b{q} \d \b{p}}{(2\pi)^{3}} \; \; p^2 \delta (\omega + E_0 - p^2/2 - V(q))
\nonumber\\
&&\times\; \rho_{0,\text{W}}(q,p,\b{q}\cdot\b{p}),
\label{}
\end{eqnarray}
which gives, after performing the integrals in spherical coordinates,
\begin{eqnarray}
\sigma^{(0)}(\omega) 
&=& \frac{4\pi}{3c\omega} \! \int_{0}^\infty \!\! \d q \; q^2 \left[ \theta(E) (2E)^{3/2}  \tilde{\rho}_{0,\text{W}}\left(q,\!\sqrt{2E} \right) \right]_{E=\omega + E_0-V(q)}\!,
\nonumber\\
\label{sigma01p}
\end{eqnarray}
where we have introduced the spherically averaged Wigner function $\tilde{\rho}_{0,\text{W}}(q,p) = \int_{-1}^{1} \d x \; \rho_{0,\text{W}}(q,p,q p x)$ and made the change of variables $E=p^2/2$ before applying the delta function. In Eq.~(\ref{sigma01p}), $\theta$ is the Heaviside step function.

\subsection{Second-order semiclassical approximation}

Using $\nabla_\b{q}^2 V(q) = V''(q) + (2/q) V'(q)$, $(\nabla_\b{q} V(q))^2 = V'(q)^2$, and $(\b{p} \cdot \nabla_\b{q})^2 V(q) = V''(q) (\b{q}\cdot \b{p}/q)^2 + V'(q) (p^2 - (\b{q}\cdot \b{p}/q)^2)/q$, we can obtain the different contributions to the second-order semiclassical correction of the photoabsorption cross section. The first contribution in Eq.~(\ref{sigma2aomega}) is
\begin{eqnarray}
\sigma^{(2a)}(\omega) &=& -\frac{\pi^2}{3c\omega} \int_{\mathbb{R}^{6}} \frac{\d \b{q} \d \b{p}}{(2\pi)^{3}} \;  \left(V''(q) + (2/q) V'(q)\right) 
\nonumber\\
&&\times\; \delta'\!\left(\omega + E_0 - p^2/2 - V(q)\right) \rho_{0,\text{W}}(q,p,\b{q}\cdot \b{p}), \;\;
\label{}
\end{eqnarray}
which gives
\begin{eqnarray}
\sigma^{(2a)}(\omega) 
                      &=& -\frac{4\pi}{12c\omega} \int_{0}^\infty \d q \;  q^2 \left(V''(q) + (2/q) V'(q)\right) 
\nonumber\\
&&\times \; \left[ \theta(E)\frac{\d}{\d E} \tilde{\rho}_{1,\text{W}}\left(q,\sqrt{2E}\right) \right]_{E=\omega + E_0 -V(q)},
\label{sigma2aomegaspher}
\end{eqnarray}
where we have introduced $\tilde{\rho}_{1,\text{W}}(q,p) = p \tilde{\rho}_{0,\text{W}}(q,p)$.

Similarly, the second contribution in Eq.~(\ref{sigma2bomega}) is
\begin{eqnarray}
\sigma^{(2b)}(\omega) &=& \frac{\pi^2}{3c\omega} 
\nonumber\\
&&\times \;
\int_{\mathbb{R}^{6}} \frac{\d \b{q} \d \b{p}}{(2\pi)^{3}} \;  \left[ V'(q)^2 - \frac{p^2}{2} \left(V''(q) + (2/q) V'(q)\right) \right] 
\nonumber\\
&& \times\; \delta''\!\left(\omega + E_0 -p^2/2 - V(q)\right) \rho_{0,\text{W}}(q,p,\b{q}\cdot \b{p}),
\label{}
\end{eqnarray}
which gives
\begin{eqnarray}
\sigma^{(2b)}(\omega) 
= \frac{4\pi}{12c\omega} \int_{0}^\infty \d q \; q^2 \Biggl[ \theta(E) \Biggl( V'(q)^2 \frac{\d^2}{\d E^2} \tilde{\rho}_{1,\text{W}}\left(q,\sqrt{2E}\right)
\nonumber\\
  - \frac{1}{2} \left(V''(q) + (2/q) V'(q)\right) \frac{\d^2}{\d E^2} \tilde{\rho}_{3,\text{W}}\left(q,\sqrt{2E}\right) \Biggl) \Biggl]_{E=\omega + E_0 -V(q)}, \;\;\;
\label{sigma2bomegaspher}
\end{eqnarray}
where we have introduced $\tilde{\rho}_{3,\text{W}}(q,p) = p^3 \tilde{\rho}_{0,\text{W}}(q,p)$.

Finally, the third contribution in Eq.~(\ref{sigma2comega}) is
\begin{eqnarray}
\sigma^{(2c)}(\omega) = \frac{\pi^2}{18c\omega} \int_{\mathbb{R}^{6}} \frac{\d \b{q} \d \b{p}}{(2\pi)^{3}} \; p^2 \phantom{xxxxxxxxxxxxxxxxxxxxxxxx}
\nonumber\\
\times\;
\left( V'(q)^2 + V''(q) (\b{q}\cdot \b{p}/q)^2 + V'(q) (p^2 - (\b{q}\cdot \b{p}/q)^2)/q \right)
\nonumber\\
 \times\; \delta'''\!\left(\omega + E_0 -p^2/2 - V(q)\right) \rho_{0,\text{W}}(q,p,\b{q}\cdot\b{p}), \;\;\;\;
\label{}
\end{eqnarray}
which gives
\begin{eqnarray}
\sigma^{(2c)}(\omega) 
 = \frac{\pi}{18c\omega} \int_{0}^\infty \d q \; q^2 \Biggl[ \theta(E) \Biggl( V'(q)^2 \frac{\d^3}{\d E^3} \tilde{\rho}_{3,\text{W}}\left(q,\sqrt{2E}\right)
\nonumber\\
+ \left( V''(q) - (1/q) V'(q) \right) \frac{\d^3}{\d E^3} \tilde{\tau}_{5,\text{W}}\left(q,\sqrt{2E}\right)
\nonumber\\
+ (1/q) V'(q) \frac{\d^3}{\d E^3} \tilde{\rho}_{5,\text{W}}\left(q,\sqrt{2E}\right)
\Biggl) \Biggl]_{E=\omega + E_0-V(q)},
\nonumber\\
\label{sigma2comegaspher}
\end{eqnarray}
where we have introduced $\tilde{\tau}_{0,\text{W}}(q,p) = \int_{-1}^1 \d x \; x^2 \rho_{0,\text{W}}(q,p,q p x)$, $\tilde{\rho}_{5,\text{W}}(q,p) = p^5 \tilde{\rho}_{0,\text{W}}(q,p)$, and $\tilde{\tau}_{5,\text{W}}(q,p) = p^5 \tilde{\tau}_{0,\text{W}}(q,p)$.

\section{Hydrogen-like atoms}
\label{sec:hydrogen}

In this section, we consider hydrogen-like atoms, i.e. one electron in the Coulomb potential $V(q) = -Z/q$. The ground state energy is $E_0=-Z^2/2$ and the ionization threshold $E_\text{thres}=0$. Beyond the ionization threshold, the photoabsorption cross section is usually called photoionization cross section.

\subsection{Expression of the photoionization cross section}

The Wigner function of the ground state of the hydrogen atom ($Z=1$) has been given in Ref.~\cite{PraMosWod-JPA-06} in the form of a one-dimensional integral that, with the help of the software Mathematica~\cite{Math12-PROG-20}, we write here as~\footnote{Note that we needed to multiply the Wigner function of Ref.~\cite{PraMosWod-JPA-06} by $4 \times (2\pi)^3$ to match the present definition.}
\begin{eqnarray}
\rho_{0,\text{W}}^{Z=1}\left(q,p,\b{q}\cdot \b{p} \right) = \int_0^1 \d u \; f(q,p,\b{q}\cdot \b{p},u),
\label{rho0WH}
\end{eqnarray}
where 
\begin{eqnarray}
f(q,p,\b{q}\cdot \b{p},u) = 16 e^{2i \b{q}\cdot \b{p} (2u-1) - 2q g(p,u)} \phantom{xxxxxxxxxxxxxxx}
\nonumber\\
\times
 \frac{(1-u)u \left( 3 + 6 q g(p,u) + 4q^2 g(p,u)^2\right)}{g(p,u)^5}, \;\;\;
\label{fRpu}
\end{eqnarray}
with $g(p,u) = \sqrt{1+4p^2 (1-u) u}$. Here, $u$ is not a physical variable but a disentanglement variable introduced to be able to perform the integration over $\b{s}$ in Eq.~(\ref{rho0W}). From this, we easily obtain the spherically averaged Wigner function as
\begin{eqnarray}
\tilde{\rho}_{0,\text{W}}^{Z=1}\left(q,p \right) = \int_0^1 \d u \; \tilde{f}(q,p,u),
\label{rho0WH}
\end{eqnarray}
where
\begin{eqnarray}
\tilde{f}(q,p,u) = 16 e^{-2 q g(p,u)} \sin \left(2 q p (2 u-1) \right) \phantom{xxxxxxxxx}
\nonumber\\
\times \frac{(1-u)u \left( 3 + 6 q g(p,u) + 4q^2 g(p,u)^2\right)}{q p (2u-1) g(p,u)^5}.
\label{fRpu}
\end{eqnarray}
The ground-state Wigner function for a hydrogen-like atom with arbitrary nuclear charge $Z$ can then be simply obtained from scaling: $\rho_{0,\text{W}}^Z\left(q,p,\b{q}\cdot \b{p} \right) = \rho_{0,\text{W}}^{Z=1} \left(Z q,p/Z,\b{q}\cdot \b{p} \right)$ and\\ $\tilde{\rho}_{0,\text{W}}^Z\left(q,p\right) = \tilde{\rho}_{0,\text{W}}^{Z=1} \left(Z q,p/Z\right)$.

The zeroth-order semiclassical photoionization cross section [Eq.~(\ref{sigma01p})] thus takes the form
\begin{eqnarray}
\sigma^{(0)}(\omega) &=& \frac{4\pi}{3c\omega} \int_0^1 \d u \int_{0}^\infty \d q \; q^2 (2(\omega + E_0+Z/q))^{3/2}  
\nonumber\\
&&\times\;\tilde{f}\left(Z q,\sqrt{2(\omega + E_0+Z/q)}/Z,u\right),
\label{sigma0H}
\end{eqnarray}
and is calculated by performing a double numerical integration over $q$ and $u$ with the software Mathematica~\cite{Math12-PROG-20}.

The terms involving the Laplacian of the Coulomb potential, $\nabla_\b{q}^2 V(q) = 4\pi Z \delta(\b{q})$, do not contribute to the second-order semiclassical correction to the photoionization cross section. Thus, the first contribution [Eq.~(\ref{sigma2aomegaspher})] vanishes
\begin{eqnarray}
\sigma^{(2a)}(\omega) &=& 0,
\label{}
\end{eqnarray}
and the second contribution [Eq.~(\ref{sigma2bomegaspher})] simplifies to, using $V'(q)=Z/q^2$,
\begin{eqnarray}
\sigma^{(2b)}(\omega) &=& \frac{\pi Z^3}{3c\omega} \int_{0}^\infty \d q \; \frac{1}{q^2}  \left[ \frac{\d^2}{\d E^2} \tilde{\rho}_{1,\text{W}}\left(Zq,\sqrt{2E}/Z\right) \right]_{E=E(\omega,q)},
\nonumber\\
\label{sigma2bH}
\end{eqnarray}
where $E(\omega,q)=\omega + E_0 +Z/q$.
Finally, using $V''(q)=-2Z/q^3$, the third contribution [Eq.~(\ref{sigma2comegaspher})] takes the form
\begin{eqnarray}
\sigma^{(2c)}(\omega) = \frac{\pi}{18c\omega} \int_{0}^\infty \d q \; \Biggl[ \frac{Z^5}{q^2} \frac{\d^3}{\d E^3} \tilde{\rho}_{3,\text{W}}\left(Zq,\sqrt{2E}/Z\right) \phantom{xxxxxxxx}
\nonumber\\
- \frac{3Z^6}{q} \frac{\d^3}{\d E^3} \tilde{\tau}_{5,\text{W}}\left(Zq,\sqrt{2E}/Z\right) \phantom{xxxxx}
\nonumber\\
+ \frac{Z^6}{q} \frac{\d^3}{\d E^3} \tilde{\rho}_{5,\text{W}}\left(Zq,\sqrt{2E}/Z\right) 
\Biggl]_{E=E(\omega,q)}. \;\;\;\;\;
\label{sigma2cH}
\end{eqnarray}
Using the software Mathematica~\cite{Math12-PROG-20}, the quantity $\tilde{\rho}_{1,\text{W}}\left(q,\sqrt{2E}\right)$, $\tilde{\rho}_{3,\text{W}}\left(q,\sqrt{2E}\right)$, $\tilde{\rho}_{5,\text{W}}\left(q,\sqrt{2E}\right)$, and $\tilde{\tau}_{5,\text{W}}\left(q,\sqrt{2E}\right)$, and their derivatives with respect to $E$ are obtained as one-dimensional integrals over $u$ similar to Eq.~(\ref{rho0WH}).

\subsection{Results and discussion}

The photoionization cross section of the hydrogen-like atom is known exactly (see, e.g., Refs.~\cite{BetSal-BOOK-57,RosVaiAstLis-MRE-20})
\begin{eqnarray}
\sigma_\text{exact}(\omega) = \frac{32 \pi^2 Z^6}{3 c \omega^4}  \frac{e^{-4 n'(\omega) \arccot n'(\omega)}}{1 - e^{-2\pi n'(\omega)}},
\label{sigmaexact}
\end{eqnarray}
where $n'(\omega)=Z/\sqrt{2(\omega+E_0)}$. 

\begin{figure}[tb]
\includegraphics[scale=0.35,angle=-90]{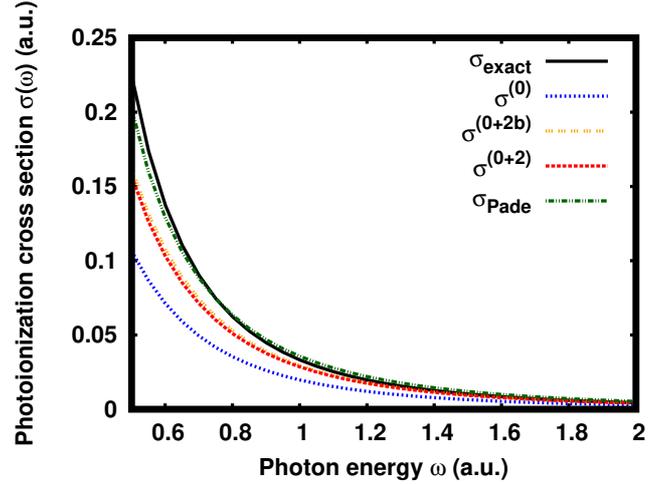}
\caption{Photoionization cross section of the hydrogen atom ($Z=1$). The exact cross section $\sigma_\text{exact}(\omega)$ [Eq.~(\ref{sigmaexact})] is compared with the zeroth-order semiclassical approximation $\sigma^{(0)}(\omega)$ [Eq.~(\ref{sigma0H})], the partial second-order semiclassical approximation $\sigma^{(0+2b)}(\omega) = \sigma^{(0)}(\omega) + \sigma^{(2b)}(\omega)$ [Eqs.~(\ref{sigma0H}) and~(\ref{sigma2bH})], the full second-order semiclassical approximation $\sigma^{(0+2)}(\omega) = \sigma^{(0)}(\omega) + \sigma^{(2b)}(\omega)+\sigma^{(2c)}(\omega)$ [Eqs.~(\ref{sigma0H}),~(\ref{sigma2bH}), and~(\ref{sigma2cH})], and the Padé approximant $\sigma_\text{Padé}(\omega)$ [Eq.~(\ref{sigmaomegapade})].
}
\label{fig:sigmavg}
\end{figure}

In Figure~\ref{fig:sigmavg}, we compare the exact cross section of the hydrogen atom ($Z=1$) with the zeroth-order semiclassical approximation $\sigma^{(0)}(\omega)$ [Eq.~(\ref{sigma0H})], the partial second-order semiclassical approximation $\sigma^{(0+2b)}(\omega) = \sigma^{(0)}(\omega) + \sigma^{(2b)}(\omega)$ [Eqs.~(\ref{sigma0H}) and~(\ref{sigma2bH})], and the full second-order semiclassical approximation $\sigma^{(0+2)}(\omega) = \sigma^{(0)}(\omega) + \sigma^{(2b)}(\omega)+\sigma^{(2c)}(\omega)$ [Eqs.~(\ref{sigma0H}),~(\ref{sigma2bH}), and~(\ref{sigma2cH})].  While the cross section $\sigma^{(0+2b)}(\omega)$ is a significant improvement over the zeroth-order cross section $\sigma^{(0)}(\omega)$, the addition of the contribution $\sigma^{(2c)}(\omega)$ has almost no effect.

As expected, the semiclassical approximation becomes more accurate as $\omega$ increases. The exact asymptotic behavior of the cross section of the hydrogen atom for large $\omega$ is~\cite{YanFaaBur-JCP-09}
\begin{eqnarray}
\sigma_\text{exact}(\omega) \isEquivTo{\omega \to \infty} \frac{16\pi \sqrt{2}}{3 c \omega^{7/2}} \approx \frac{0.172}{\omega^{7/2}}.
\label{sigmaomegainf}
\end{eqnarray}
Numerically, we find $\lim_{\omega \to \infty} \omega^{7/2} \sigma^{(0)}(\omega) \approx 0.11$ and\\ $\lim_{\omega \to \infty} \omega^{7/2} \sigma^{(0+2)}(\omega)  \approx 0.17$. So, $\sigma^{(0)}(\omega)$ has the correct behavior in $1/\omega^{7/2}$ but not the correct prefactor, while $\sigma^{(0+2)}(\omega)$ has the correct prefactor.

To improve the accuracy at small $\omega$, one may resum the semiclassical expansion in Eq.~(\ref{sigmaomegaexpand}) using a $[0/1]$ Padé approximant
\begin{eqnarray}
\sigma_\text{Padé}(\omega) = \frac{\sigma^{(0)}(\omega)}{ 1 - \sigma^{(2)}(\omega) / \sigma^{(0)}(\omega) }, 
\label{sigmaomegapade}
\end{eqnarray}
which is also plotted in Figure~\ref{fig:sigmavg}. We see that the Padé approximant is quite effective indeed to improve the accuracy at small $\omega$.

\begin{figure}[tb]
\includegraphics[scale=0.35,angle=-90]{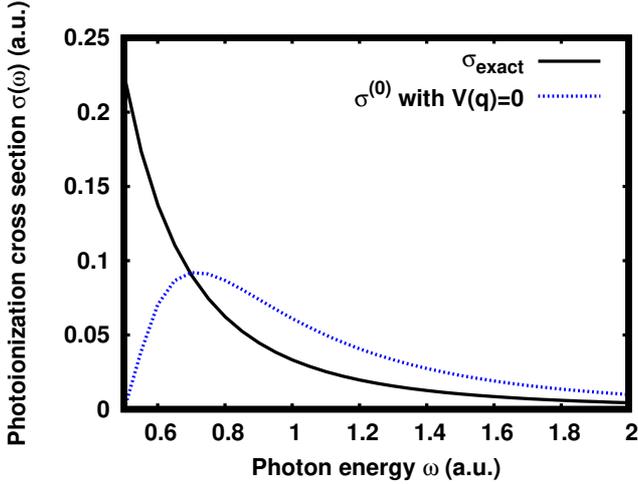}
\caption{Photoionization cross section of the hydrogen atom. The exact cross section $\sigma_\text{exact}(\omega)$ [Eq.~(\ref{sigmaexact})] is compared with the zeroth-order semiclassical approximation $\sigma^{(0)}(\omega)$ where the Coulomb potential $V(q)$ has been neglected in $B_\text{W}(\b{q},\b{p})$.
}
\label{fig:hvelocitynoV}
\end{figure}

Finally, Figure~\ref{fig:hvelocitynoV} shows the effect of neglecting the Coulomb potential $V(q)$ in $B_\text{W}(\b{q},\b{p})$, equivalent to using the free-particle plane-wave continuum states. This changes completely the shape of the spectrum. In particular, the cross section is now zero at the ionization threshold, in accordance with the Wigner-threshold law~\cite{Wig-PR-48,SadBohCavEsrFabMacRau-JPB-00} which predicts this behavior for potentials lacking a long-range attractive $-1/q$ Coulomb tail. The obtained spectrum has in fact a similar shape as the one obtained in Hartree-Fock~\cite{ZapLupTou-JCP-19} whose continuum states only see an exponentially decaying effective potential.

\section{Helium-like atoms}
\label{sec:helium}

In this section, we consider helium-like atoms, i.e. $N=2$ electrons. The phase-space variables are now $\b{q} \equiv (\b{q}_1,\b{q}_2) \in \mathbb{R}^6$ and $\b{p} \equiv (\b{p}_1,\b{p}_2) \in \mathbb{R}^6$, and the classical Hamiltonian is
\begin{eqnarray}
H(\b{q},\b{p}) =\frac{p_1^2}{2} + \frac{p_2^2}{2}+ V(\b{q}_1,\b{q}_2),
\end{eqnarray}
where $p_i=||\b{p}_i||$ and $q_i=||\b{q}_i||$, and the potential is $V(\b{q}_1,\b{q}_2)=-Z/q_1-Z/q_2 +1/||\b{q}_1-\b{q}_2||$.

\subsection{Wigner function of the ground state}

As an approximation to the exact ground-state wave function, we consider the Hartree-Fock (HF) wave function, $\Phi(\b{q}_1,\b{q}_2) = \phi(\b{q}_1) \phi(\b{q}_2)$, where $\phi$ is the HF 1s orbital. The associated Wigner function can be factorized as
\begin{eqnarray}
\rho_{\text{HF},\text{W}}(\b{q}_1,\b{q}_2,\b{p}_1,\b{p}_2) = \rho_{\phi,\text{W}}(q_1,p_1,\b{q}_1 \cdot \b{p}_1) \rho_{\phi,\text{W}}(q_2,p_2,\b{q}_2 \cdot \b{p}_2),
\nonumber\\
\label{rhoHFW}
\end{eqnarray}
where $\rho_{\phi,\text{W}}(q,p,\b{q}\cdot \b{p})$ is the Wigner function associated with the 1s orbital $\phi$
\begin{eqnarray}
\rho_{\phi,\text{W}}(q,p,\b{q}\cdot \b{p}) =  \int_{\mathbb{R}^{3}} \d \b{s} \; e^{-i \b{p} \cdot \b{s}} \phi(\b{q}-\b{s}/2) \phi(\b{q}+\b{s}/2).
\label{}
\end{eqnarray}
As usual in quantum chemistry, the orbital $\phi$ is expanded on $M$ Gaussian basis functions $\chi_i(\b{q})= (2\alpha_i/\pi)^{3/4} e^{-\alpha_i q^2}$, where $\alpha_i$ are fixed exponents,
\begin{eqnarray}
\phi(\b{q}) = \sum_{i=1}^M c_i \chi_i (\b{q}),
\end{eqnarray}
and $c_i$ are coefficients found by solving the HF self-consistent equation. Following Ref.~\cite{DahSpr-MP-82}, the corresponding Wigner function is easily obtained as 
\begin{eqnarray}
\rho_{\phi,\text{W}}(q,p,\b{q}\cdot \b{p}) = \sum_{i=1}^M \sum_{j=1}^M c_i c_j P_{i,j}(q,p,\b{q}\cdot \b{p}),
\end{eqnarray}
where
\begin{eqnarray}
P_{i,j}(q,p,\b{q}\cdot \b{p}) &=&  \int_{\mathbb{R}^{3}} \d \b{s} \; e^{-i \b{p} \cdot \b{s}} \chi_i(\b{q}-\b{s}/2) \chi_j(\b{q}+\b{s}/2).
\nonumber\\
&=& 2^3 \left( \beta_{i,j} \gamma_{i,j}\right)^{3/4} e^{-\gamma_{i,j} q^2} e^{-\beta_{i,j} p^2} e^{-2 \i \tau_{i,j} \b{q} \cdot \b{p}},
\label{}
\end{eqnarray}
with $\beta_{i,j} = 1/(\alpha_i+\alpha_j)$, $\gamma_{i,j}=4 \alpha_i\alpha_j \beta_{i,j}$, $\tau_{i,j}=(\alpha_i-\alpha_j) \beta_{i,j}$. The Wigner function can be rewritten as
\begin{eqnarray}
\rho_{\phi,\text{W}}(q,p,\b{q}\cdot \b{p}) = \sum_{i=1}^M c_i^2 f_{i,i}(q,p) + \sum_{i=1}^M \sum_{j=i+1}^M c_i c_j f_{i,j}(q,p,\b{q}\cdot \b{p}),
\nonumber\\
\end{eqnarray}
where 
\begin{eqnarray}
f_{i,i}(q,p)= P_{i,i}(q,p,\b{q}\cdot \b{p})=2^3 e^{-2\alpha_i q^2} e^{-p^2/(2\alpha_i)},
\end{eqnarray}
and 
\begin{eqnarray}
f_{i,j}(q,p,\b{q}\cdot \b{p}) &=& P_{i,j}(q,p,\b{q}\cdot \b{p})+P_{j,i}(q,p,\b{q}\cdot \b{p})
\nonumber\\
&=&2^4 ( \beta_{i,j} \gamma_{i,j})^{3/4} e^{-\gamma_{i,j} q^2} e^{-\beta_{i,j} p^2} \cos(2 \tau_{i,j} \b{q} \cdot \b{p}).
\nonumber\\
\end{eqnarray}

\subsection{Expression of the photoionization cross section}

\subsubsection{Zeroth-order contribution}

Using the HF Wigner function in Eq.~(\ref{rhoHFW}) and the corresponding exact HF ground-state energy $E_0^\text{HF}=-2.861680$ a.u. (giving an ionization threshold of $E_\text{thres}-E_0^\text{HF}=0.861680$ a.u.), the zeroth-order photoionization cross section in Eq.~(\ref{sigma0omega}) becomes
\begin{eqnarray}
\sigma^{(0)}(\omega) = \frac{4\pi^2}{3c\omega} \int_{\mathbb{R}^{12}} \frac{\d \b{q}_1 \d \b{q}_2 \d \b{p}_1 \d \b{p}_2}{(2\pi)^{6}} \; (p_1^2 + p_2^2 + 2 \b{p}_1 \cdot \b{p}_2) 
\nonumber\\
\phantom{}\times 
 \delta(\omega + E_0^\text{HF} -p_1^2/2-p_2^2/2 - V(\b{q}_1,\b{q}_2))
\nonumber\\
\phantom{x}\times 
\rho_{\phi,\text{W}}(q_1,p_1,\b{q}_1 \cdot \b{p}_1) \rho_{\phi,\text{W}}(q_2,p_2,\b{q}_2 \cdot \b{p}_2). \;
\label{}
\end{eqnarray}
Using spherical coordinates for $\b{p}_1$ and $\b{p}_2$, and integrating over the angles, we get
\begin{eqnarray}
\sigma^{(0)}(\omega) = \frac{4\pi^2}{3c\omega (2\pi)^{4}} 
\int_{\mathbb{R}^{6}}\!\! \d \b{q}_1 \d \b{q}_2 \int_0^\infty \!\! \d p_1 \int_0^\infty \!\! \d p_2 p_1^2 p_2^2 (p_1^2 + p_2^2)  
\nonumber\\
\times \;\delta(\omega + E_0^\text{HF} -p_1^2/2-p_2^2/2 - V(\b{q}_1,\b{q}_2))
\nonumber\\
\times  \;
\tilde{\rho}_{\phi,\text{W}}(q_1,p_1) \tilde{\rho}_{\phi,\text{W}}(q_2,p_2). \phantom{xxxxxxxxxxx}
\label{}
\end{eqnarray}
where $\tilde{\rho}_{\phi,\text{W}}(q,p) = \int_{-1}^{1} \d x \rho_{\phi,\text{W}}(q,p,q p x)$, and we have used that the fact the integral over the angles of the term involving $\b{p}_1 \cdot \b{p}_2$ vanishes because $\rho_{\phi,\text{W}}(q,p,q p x)$ is an even function of $x$. Using polar coordinates $p_1= \eta \cos \varphi$ and $p_2= \eta \sin \varphi$, and making the change of variables $E=\eta^2/2$ before applying the delta function, we find
\begin{eqnarray}
\sigma^{(0)}(\omega) = \phantom{xxxxxxxxxxxxxxxxxxxxxxxxxxxxxxxxxxxxxxxxx}
\nonumber\\
\frac{2}{3c\omega} 
\int_0^\infty \!\! \d q_1  \int_0^\infty \!\! \d q_2 \int_{-1}^{1} \!\! \d x \int_0^{\pi/2} \!\! \d \varphi  q_1^2 q_2^2 (\cos \varphi)^2 (\sin \varphi)^2 \phantom{xxxxx}
\nonumber\\
\times \Biggl[ \theta(E) (2E)^3 \tilde{\rho}_{\phi,\text{W}}(q_1,\!\sqrt{2E} \cos \varphi) \phantom{xxxxxxxxx}
\nonumber\\
\times \tilde{\rho}_{\phi,\text{W}}(q_2,\!\sqrt{2E} \sin \varphi) \Biggl]_{E=E(\omega,q_1,q_2,x)}, \phantom{xxxx}
\label{sigma0He}
\end{eqnarray}
where 
\begin{eqnarray}
E(\omega,q_1,q_2,x) = \omega + E_0^\text{HF} +\frac{Z}{q_1} +\frac{Z}{q_2} -\frac{1}{\sqrt{q_1^2+q_2^2 -2 q_1 q_2 x}}.
\nonumber\\
\end{eqnarray}

\subsubsection{Second-order contribution}

Similarly to hydrogen-like atoms, the Laplacian-like terms,\\ $\b{D}^2 V(\b{q})$ and $\nabla_\b{q}^2 V(\b{q})$, should not contribute to the second-order photoionization cross section. Thus, $\sigma^{(2a)}(\omega)$ [Eq.~(\ref{sigma2aomega})] is zero and $\sigma^{(2b)}(\omega)$ [Eq.~(\ref{sigma2bomega})] simplifies to
\begin{eqnarray}
\sigma^{(2b)}(\omega) = \frac{\pi^2}{3c\omega} \int_{\mathbb{R}^{12}} \frac{\d \b{q}_1 \d \b{q}_2 \d \b{p}_1 \d \b{p}_2 }{(2\pi)^{6}} \; \left( \b{D} V(\b{q}_1,\b{q}_2) \right)^2 
\nonumber\\
\times \delta''\!(\omega + E_0^\text{HF} -p_1^2/2-p_2^2/2 - V(\b{q}_1,\b{q}_2))
\nonumber\\
\times 
\rho_{\phi,\text{W}}(q_1,p_1,\b{q}_1 \cdot \b{p}_1) \rho_{\phi,\text{W}}(q_2,p_2,\b{q}_2 \cdot \b{p}_2),
\label{}
\end{eqnarray}
where $\b{D} V(\b{q}_1,\b{q}_2) = Z(\b{q}_1/q_1^3 + \b{q}_2/q_2^3)$. Using spherical coordinates for $\b{p}_1$ and $\b{p}_2$, we get
\begin{eqnarray}
\sigma^{(2b)}(\omega) = \frac{\pi^2}{3c\omega (2\pi)^{4}} \int_{\mathbb{R}^{6}} \!\! \d \b{q}_1 \d \b{q}_2 \int_0^\infty \!\! \d p_1 \int_0^\infty \!\! \d p_2 \; p_1^2 p_2^2 \phantom{xxx}
\nonumber\\
\times \left( \b{D} V(\b{q}_1,\b{q}_2) \right)^2 
\delta''\!(\omega + E_0^\text{HF} -p_1^2/2-p_2^2/2 - V(\b{q}_1,\b{q}_2))
\nonumber\\
\times \tilde{\rho}_{\phi,\text{W}}(q_1,p_1) \tilde{\rho}_{\phi,\text{W}}(q_2,p_2). \;\;
\label{}
\end{eqnarray}
Using now polar coordinates $p_1= \eta \cos \varphi$ and $p_2= \eta \sin \varphi$, and making the change of variables $E=\eta^2/2$, we find
\begin{eqnarray}
\sigma^{(2b)}(\omega) = 
\frac{Z^2}{6c\omega} \int_0^\infty \!\! \d q_1 \int_0^\infty \!\! \d q_2 \int_{-1}^{1} \!\! \d x\int_0^{\pi/2} \!\!\d \varphi \;  (\cos \varphi)^2 (\sin \varphi)^2 
\nonumber\\
\times \left( \frac{q_1^2}{q_2^2} + \frac{q_2^2}{q_1^2} + 2x\right) 
\Biggl[ \theta(E) \frac{\d^2}{\d E^2}\Bigl( (2E)^{2}
\tilde{\rho}_{\phi,\text{W}}(q_1,\sqrt{2E} \cos \varphi) 
\nonumber\\
\times \tilde{\rho}_{\phi,\text{W}}(q_2, \sqrt{2E} \sin \varphi) \Bigl) \Biggl]_
{E=E(\omega,q_1,q_2,x)}. \;\;\; \;\;\;
\label{sigma2bHe}
\end{eqnarray}

Based on the results obtained for the hydrogen atom, we expect the last second-order contribution $\sigma^{(2c)}(\omega)$ [Eq.~(\ref{sigma2comega})] to be small and so we will not attempt to calculate it.

\subsection{Results and discussion}

\begin{figure}[tb]
\includegraphics[scale=0.35,angle=-90]{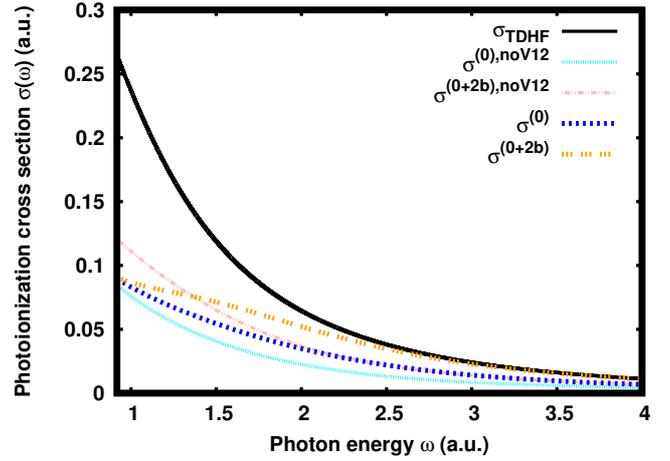}
\caption{Photoionization cross section of the helium atom ($Z=2$). The reference TDHF cross section $\sigma_\text{TDHF}(\omega)$~\cite{ZapLupTou-JCP-19,SchZapLevCanLupTou-JCP-22} is compared with the zeroth-order semiclassical cross section $\sigma^{(0)}(\omega)$ [Eq.~(\ref{sigma0He})] and the partial second-order semiclassical cross sections $\sigma^{(0+2b)}(\omega) = \sigma^{(0)}(\omega) + \sigma^{(2b)}(\omega)$ [Eqs.~(\ref{sigma0He}) and~(\ref{sigma2bHe})]. Also reported are the approximate zeroth-order and partial second-order semiclassical cross sections $\sigma^{(0),\text{noV12}}(\omega)$ and $\sigma^{(0+2b),\text{noV12}}(\omega) = \sigma^{(0),\text{noV12}}(\omega) + \sigma^{(2b),\text{noV12}}(\omega)$ [Eqs.~(\ref{sigma0He}) and~(\ref{sigma2bHe})] in which the two-electron Coulomb interaction $V_{12}(\b{q}_1,\b{q}_2)=1/||\b{q}_1-\b{q}_2||$ has been neglected.
}
\label{fig:he_sigma}
\end{figure}

We performed a HF calculation with the uncontracted Gaussian cc-pVDZ basis set~\cite{WooDun-JCP-94,PriAltDidGibWin-JCIM-19} (containing $M=4$ s Gaussian basis functions) to obtain the 1s HF occupied orbital of the helium atom ($Z=2$). The exponents $\alpha_i$ are $38.36$, $5.77$, $1.24$, $0.2976$, and the corresponding coefficients $c_i$ are $0.02380882$, $0.15489122$, $0.46998667$, $0.51302690$. We then calculated by numerical integration with the software Mathematica~\cite{Math12-PROG-20} the zeroth-order semiclassical cross section $\sigma^{(0)}(\omega)$ [Eq.~(\ref{sigma0He})] and the partial second-order correction $\sigma^{(2b)}(\omega)$ [Eq.~(\ref{sigma2bHe})]. The numerical integration for $\sigma^{(2b)}(\omega)$ turned out to be delicate and requires a somewhat costly local-adaptive algorithm. For this reason, we also considered the approximation (refer to as ``noV12'') consisting in neglecting the two-electron Coulomb interaction $V_{12}(\b{q}_1,\b{q}_2)=1/||\b{q}_1-\b{q}_2||$ in the expression of $V(\b{q}_1,\b{q}_2)$. This is done simply by replacing $E(\omega,q_1,q_2,x)$ in Eqs.~(\ref{sigma0He}) and~(\ref{sigma2bHe}) by $E^\text{noV12}(\omega,q_1,q_2) = \omega + E_0^\text{HF} +Z/q_1 +Z/q_2$. This eliminates the numerical integration over the variable $x$ and makes the remaining numerical integration easier to perform. The resulting zeroth-order and partial second-order semiclassical cross sections are designated by $\sigma^{(0),\text{noV12}}(\omega)$ and $\sigma^{(0+2b),\text{noV12}}(\omega)$, respectively.

Figure~\ref{fig:he_sigma} reports these photoionization cross sections for the helium atom. As reference, we use the linear-response time-dependent Hartree-Fock (TDHF) cross section calculated with a B-spline basis set~\cite{ZapLupTou-JCP-19,SchZapLevCanLupTou-JCP-22}. Similarly to the case of the hydrogen atom, the zeroth-order cross section $\sigma^{(0)}(\omega)$ is always too small but improves at $\omega$ increases. The partial second-order cross section $\sigma^{(0+2b)}(\omega)$ constitutes an improvement over $\sigma^{(0)}(\omega)$ and is accurate at high energy. The approximate partial second-order cross section $\sigma^{(0+2b),\text{noV12}}(\omega)$ is significantly less accurate at high energy, showing that one should avoid neglecting the two-electron Coulomb interaction.

\section{Conclusions and future directions}
\label{sec:conclusion}

In this work, we have developed semiclassical approximations for calculating photoabsorption/photoionization cross sections. The approximations only require to have the Wigner function of the ground state and bypass the need to explicitly calculate the continuum states. Examples in electronic-structure theory on the hydrogen and helium atoms suggest that these approximations can be used to obtain good estimates of photoabsorption/photoionization cross sections at high energy. 

However, at least two limitations remain. First, we do not have any a priori estimates of the errors made by these semiclassical approximations. Second, it seems difficult to extend the present calculations done by deterministic numerical integration to a larger number of particles. Regarding the latter point, a possible strategy to treat general systems would be to calculate the $6N$-dimensional phase-space integrals by Monte Carlo sampling of the Wigner function~\cite{KubLasWeb-JCP-09}. The functions to average in Eq.~(\ref{sigma0omega}) and in Eqs.~(\ref{sigma2aomega})-(\ref{sigma2comega}) are singular (they contain the Dirac delta function and its derivatives) but the Monte Carlo techniques developed in Refs.~\cite{AssCafSce-PRE-07,TouAssUmr-JCP-07,BorAssRotVui-MP-13} could be used to efficiently calculate these averages.

More generally, the type of Wigner-based semiclassical approximations developed in the present work could be useful in quantum many-body theory to calculate efficiently the contribution of the high-lying continuum states to various quantities such as second-order or coupled-cluster correlation energies~\cite{SzaOst-BOOK-96,HelJorOls-BOOK-02} which are known to converge slowly with the size of the one-particle basis set for a two-particle interaction with a hard short-range part such as the Coulomb interaction~\cite{HelKloKocNog-JCP-97,HalHelJorKloKocOlsWil-CPL-98}.


\end{document}